\documentclass{iau} 
\usepackage{graphicx}

\def  \lagn     {$L_{\rm AGN}$}

\def  \lirsf     {$L_{\rm IR, SF}$}
\def\sf{star formation}
\def \mbh {M_{\bullet}}
\def \apj {{\it ApJ}}
\def \apjl {{\it ApJL}}
\def \apjs {{\it ApJS}}
\def \aj {{\it AJ}}
\def \mnras {{\it MNRAS}}
\def \araa {{\it ARA\&A}}
\def \aap {{\it A\&A}}
\def \nat {{\it Nature}}

\title[AGN-host connection] 
{Tips learned from panchromatic modeling of AGNs}

\author[Y.Sophia Dai]   
{Y.Sophia Dai$^1$}

\affiliation{$^1$CASSACA, National Astronomical Observatories of China (NAOC)
\\ email: {\tt daysophia@gmail.com} \\[\affilskip]}

\pubyear{2018}
\volume{341}  
\jname{Challenges in Panchromatic Modelling with Next Generation Facilities}
\editors{M. Boquien, E. Lusso, C. Gruppioni, \& P. Tissera}
\begin{document}

\maketitle

\begin{abstract}
I will review the tips learned from panchromatic modeling of active galactic nuclei (AGNs), based on our recent work to study the relationship between AGN and star formation (SF). Several AGN SED models are compared, and significant AGN contribution is found in the IR luminosities and corresponding star formation rate (SFR). I will review the AGN-SF relation and how different parameters and sample selections affect the observed correlation.  I will then report on the constant ratio discovered between the SFR and the black hole mass accretion rate (BHAR), and their implications on the gas supply and galaxy formation history of these systems. Caveats and important questions to answer are summarized at the end.
\end{abstract}

\firstsection 
\vspace{-0.6cm}
\section{Introduction}
Twenty years have passed since the empirical connections found 
between the supermassive black hole (SMBH) mass ($\mbh$) and properties of their host galaxies,
including the stellar velocity dispersion (M-$\sigma$), 
the bulge luminosity, and the bulge mass by e.g. \cite{magorrian98, mnh03}. 
The stellar mass-$\mbh$ ratio shows a larger variation than the bulge mass-$\mbh$ ratios,
which ranges from a few hundreds to a few thousands. 
This possibly indicates intrinsically different relations for different galaxy types  (\cite{mnh03, knh13, rnv15}). 

Given their significantly different sizes and masses,
the finding of these correlations 
triggered the search for the intrinsic physical drive that connects SMBHs and galaxies. 
Various evolution models have been proposed, 
including the merger theory, e.g., \cite{hopkins06}, 
the cold-flow scheme, e.g., \cite{springel05}, 
or purely mathematical model, e.g., \cite{peng07}. 
In these models, AGN feedback is often needed to modulate the process,
locally or globally, in forms of winds, jets or radiation perturbations, e.g., \cite{fabian12}. 

In practice, statistical samples of AGNs or star forming galaxies are used
to study the AGN-SF relation.  
Typical sample selection utilizes the fact the AGN and SF dominate different parts of the spectrum.  
AGNs are often selected in the X-ray, e.g., \cite{lutz10, mullaney12, dai18}, 
or by highly ionized and often wide optical lines, e.g., \cite{netzer09, matsuoka15,harris16}, 
while IR luminosity or certain lines (e.g H$\alpha$) is chosen to represent SF.
Since AGN selections in the X-ray and optical are not targeting the same nucleus regions,
and optical lines and infrared luminosities are not tracing the same SF regions (gas $vs$ cold dust), 
direct comparisons could be confusing. 
Recently, \cite{dai18} constructed a sample of IR-bright AGNs,
with detections in both the X-ray and far-IR, 
aiming to focus on the phase where both BH accretion and SF are active. 
\begin{figure*}[b]
\vspace{-0.4cm}
\begin{center}
 \includegraphics[width=3.5in]{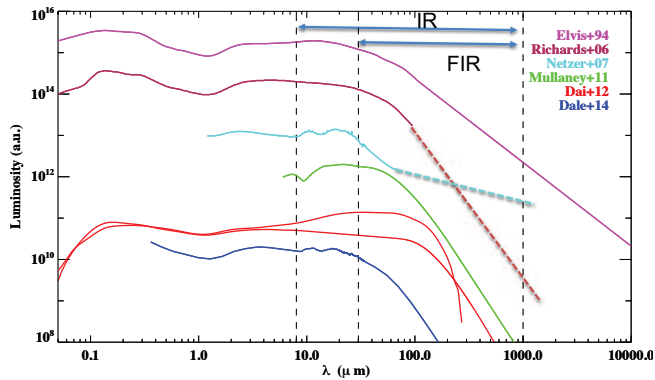} 
 \caption{Comparison of AGN SEDs from the literature in arbitrary unit for display purpose. 
 General consistency in the total IR luminosity and relatively larger variation in the far-IR luminosity 
 are found (Sec~\ref{sec:sed}).}
   \label{fig:sedcf}
   \vspace{-0.2cm}
\end{center}
\end{figure*}

\vspace{-0.5cm}
\section{SED Analysis}
\label{sec:sed}
SED decomposition is one of the most common practices
to derive the relative luminosities of AGN and SF components.
This is typically achieved by fitting models, 
either using existing de-composition codes (e.g. CIGALE, MagPhys, GRASIL), 
or users' own decomposition templates, e.g. \cite{rosario18, rivera16}. 
Despite large scatter, the AGN mean SEDs have shown 
surprising uniformity over $z$, luminosity, and Eddington ratios (e.g., \cite{elvis94, richards06, dai12}),
with a big-blue-bump in the UV-optical, a near-IR bump,
followed by a not-so-well-constrained far-IR decline due to lack of observation data. 
With $Herschel$, a separation between the younger, far-IR bright population
and the older, far-IR faint AGN population has been reported by e.g.,  \cite{dai12}. 

In Fig.~\ref{fig:sedcf}, 
we compare the AGN IR SED models from \cite{elvis94, richards06, mullaney11, dai12}, and \cite{dale14}. 
A general consistency ($\sim$ 0.2\,dex) is found for the integrated 
IR luminosities (8-1000$\mu$m, normalized at 6 $\mu$m)
amongst different models. 
Counter-intuitively, after removing the near- to mid- IR---believed to be dominated by AGN 
thus affected by AGN variability, 
a larger discrepancy ($\sim$0.6\,dex) is found in the far-IR (30-1000$\mu$m).  
The intrinsic variation of the AGN SED in the far-IR,
and the interpolation for some of the templates,
could both contribute to this far-IR inconsistency. 
The total variation, though, is still $\leq$ 1 dex. 
To better use the known information in the X-ray,
we developed a 3-step AGN IR decomposition method as described in \cite{dai18}. 
Average AGN contributions of 23\% and 11\% are found for the total and far IR luminosities, respectively. 
We conclude that AGN removal is essential, but uncertain, in the IR. 
\vspace{-0.1cm}
\begin{figure*}[b]
\begin{center}
\vspace{-0.4cm}
 \includegraphics[width=3.5in]{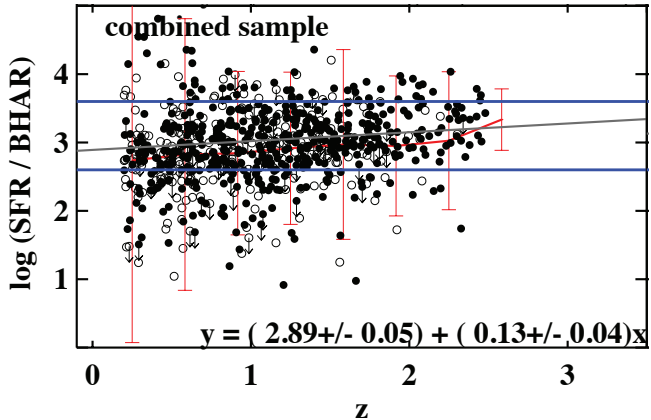} 
 \vspace{-0.25cm}
 \caption{The SFR/BHAR ratio derived for the IR-bright AGNs from \cite{dai18}. 
 An overall flat trend is observed with a median and deviation of log(SFR/BHAR)$= 2.9\,\pm\,0.5$. 
 The two lines mark the $M_*/\mbh$ ratios of 2.6 and 3.6,
 as reported in \cite{magorrian98} and \cite{rnv15}, respectively.} 
   \label{fig:mdot}
\end{center}
\end{figure*}

\vspace{-0.5cm}
\section{AGN-SF correlation in two formats}
{\it 3.1. $L_{\rm SFR}$/$L_{\rm AGN}$ ratio}

Observationally, various, sometimes contradictory correlations have been reported between 
AGN and SF luminosities,
be it suppressed SFR in luminous AGN host by e.g., \cite{page12, barger15};
or flat or unrelated AGN-SF luminosities by e.g., \cite{stanley15});
or bi-modality and overall linear correlations, often with a sample-dependent correlation coefficient,
by e.g. \cite{lutz10, harris16, pitchford16, shimizu17, dai18}. 
We note that AGN populations may be intrinsically different at the X-ray luminous and faint ends, 
as indicated by observations of X-ray bright sources with ALMA and SCUBA2,
also indicated by the bi-modality mentioned above.
Therefore, one has to be careful in interpreting the observed trend when stacking in the X-ray or IR is used. 

{\it 3.2. SFR/BHAR ratio}

The radio between BHAR and SFR
can be used to avoid a false correlation due to $z$ effect. 
Although with a scatter of $\sim$0.5\,dex, we found a constant SFR/BHAR ratio over $z$, $\mbh$, and ${M_*}$, 
with the $\dot{M}$ ratio mostly in the log (SFR/BHAR)$=$ 2.6-3.6 range (Fig.~\ref{fig:mdot}). 
This is consistent with the scenario that on an average basis, 
the galaxy and the black hole form at a fixed rate similar to the locally observed mass ratio. 
Though with large error bars, 
some recent studies show a $M_*$ dependence of this $\dot{M}$ ratio, e.g., \cite{yang17, cowley18}, 
while others find a constant ratio independent of $M_*$, e.g. \cite{mullaney12}, Dai et al., (in prep). 

{\it 3.3. Caveats of the correlation studies}

$\bullet$ Malmquist bias is prevalent, which could result in false increase at the luminous end.

\begin{figure}[b]
\begin{center}
\vspace{-0.4cm}
 \includegraphics[width=3.5in]{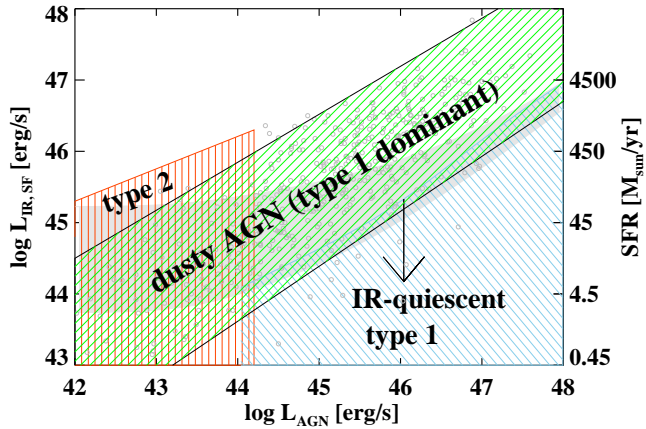} 
 \caption{A toy-model illustration of the locations of different AGN populations in the \lagn-\lirsf\ plane.
Sample selections can significantly affect the observed trend.
This model assumes that the AGN-SF correlation are most significant during a phase
where both AGN and SF are active (green region). 
Grey circles mark the \cite{dai18} sample.
Stacking of the IR-undetected X-ray sources (blue region),
or of the X-ray undetected IR sources (red region), 
could both flatten the observed relation.}
   \label{fig:types}
\end{center}
\vspace{-0.4cm}
\end{figure}

$\bullet$ Sample-dependence is important. 
Fig.~\ref{fig:types} illustrates a simplified picture assuming that the AGN-SF correlation only exists
during the active phase, during which 
AGNs coexist with active star formation, resulting in the observed fixed-fraction of gas inflow. 
This explains the flattening slopes found in samples with 
higher fraction of obscured objects, e.g. recovered by stacks in the X-ray. 

$\bullet$ The choice of binning results in different correlation slopes, 
or the lack of any correlation. This can be due to 1) the different variability time scale for AGNs and SF; 
2) purely mathematical effects due to choices of the dependent parameters (\cite{rnv15, dai18}). 

\vspace{-0.6cm}
\section{Summary}
Based on SED analysis of the IR-bright AGN sample in \cite{dai18}, we found:
\begin{enumerate}
\item 
\lagn-\lirsf, and SFR-BHAR correlations have been confirmed by various observational studies, 
consistent with the scenario of a common gas/mass supply for SMBH and the host galaxy. 

\item
A nearly constant ratio of log(SFR/BHAR) $\sim$ 2.9 is observed, agreeing with the local stellar mass $vs$ SMBH mass ratios,
indicating homogeneous evolution across $z$. The effect of ${M_*}$ is still debatable and under investigation.

\item
Several caveats can potentially mask out an intrinsic AGN-SF correlations, e.g. 
1). intrinsic scatter and uncertainties ($\sim$0.5\,dex) due to various SFR \& BHAR estimators;  
2). Selection bias and Malmquist bias;
3). Mixing of different populations by stacking and sample selection;
4). Binning method (variability, choice of free parameter).
\end{enumerate}

Future studies need to construct a clean AGN sample, 
of similar physical properties, at similar evolutionary stages, 
to study the intrinsic AGN SED(s), and to explain the flat SFR/BHAR ratios, their dependence, and their indication on the evolution history.

\begin{discussion}

\discuss{David Rosario}{Existing work on panchromatic modelling (e.g., Rosario$+$2018) do include x-ray constraints. An important bias is that luminous AGN tend to be in more massive hosts, which, regardless of AGN, do show higher SFRs. How does this affect $L_{\rm AGN}-L_{\rm SFR}$ relationships?}

\discuss{Dai Y.Sophia}{Good point. The higher AGN fraction at the luminous massive end is likely to contribute to the increase in SFR in general. Thus it is helpful to make the comparison in a mass normalized fashion, i.e. $L_{\rm AGN}$/${M_*}  vs$ SFR, or $L_{\rm AGN}/M_{\rm BH}$ vs SFR/$M_*$, or $L_{\rm AGN}/M_{\rm BH}$ vs SFR. It would be interesting to see how these parameters look in large samples. A possible bias would be the uncertainties in the mass estimates.}

\end{discussion}

\end{document}